\documentclass[aps,superscriptaddress,showpacs,preprint,pra]{revtex4-1} 
\usepackage{epsfig}
\usepackage{amsfonts}
\usepackage{amsmath}
\usepackage{xcolor}
\usepackage[normalem]{ulem}
\usepackage[bookmarks=true]{hyperref}

\begin{document}

\title{Detecting dimensional crossover and finite Hilbert space through entanglement entropies.}

\author{Mariano Garagiola}
\email{mgaragiola@famaf.unc.edu.ar}
\affiliation{Facultad de Matem\'atica, Astronom\'{\i}a y F\'{\i}sica,
Universidad Nacional de C\'ordoba and IFEG-CONICET, Ciudad Universitaria,
X5000HUA C\'ordoba, Argentina}

\author{Eloisa Cuestas}
\email{mecuestas@famaf.unc.edu.ar}
\affiliation{Facultad de Matem\'atica, Astronom\'{\i}a y F\'{\i}sica,
Universidad Nacional de C\'ordoba and IFEG-CONICET, Ciudad Universitaria,
X5000HUA C\'ordoba, Argentina}

\author{Federico~M.~Pont}
\email{pont@famaf.unc.edu.ar}
\affiliation{Facultad de Matem\'atica, Astronom\'{\i}a y F\'{\i}sica,
Universidad Nacional de C\'ordoba and IFEG-CONICET, Ciudad Universitaria,
X5000HUA C\'ordoba, Argentina}

\author{Pablo Serra}
\email{serra@famaf.unc.edu.ar}
\affiliation{Facultad de Matem\'atica, Astronom\'{\i}a y F\'{\i}sica,
Universidad Nacional de C\'ordoba and IFEG-CONICET, Ciudad Universitaria,
X5000HUA C\'ordoba, Argentina}

\author{Omar Osenda}
\email{osenda@famaf.unc.edu.ar}
\affiliation{Facultad de Matem\'atica, Astronom\'{\i}a y F\'{\i}sica,
Universidad Nacional de C\'ordoba and IFEG-CONICET, Ciudad Universitaria,
X5000HUA C\'ordoba, Argentina}

\begin{abstract}
The information content of the two-particle one- and two-dimensional Calogero model is studied using
the von Neumann and R\'enyi entropies. The one-dimensional model is shown to have non-monotonic entropies with finite values in the large interaction strength limit.
On the other hand, the von Neumann entropy of the two-dimensional model with
isotropic confinement is a monotone increasing function of the interaction strength which diverges logarithmically. By considering an anisotropic confinement in the two-dimensional case we show that the one-dimensional behavior is eventually reached when the anisotropy increases. The crossover from two to one dimensions is demonstrated using the harmonic approximation and it is shown that the von Neumann divergence only occurs in the isotropic case. The R\'enyi entropies are used to highlight the structure of the model spectrum. In  particular, it is shown that these
entropies have a non-monotonic and non-analytical behavior in the neighborhood of the interaction strength parameter values where the Hilbert space and, consequently,
the spectrum of the reduced density matrix are both finite.
\end{abstract}
\date{\today}

\maketitle


\section{Introduction}

In the last years there has been a growing interest in models of many 
interacting 
particles with continuous variables \cite{Killoran2014,Iemini2013}. This 
interest is twofold fueled by, on one hand, some unexpected 
physical traits shown by the models and, on the other hand, that they can be 
analytically treated to a 
great extent providing exact solutions based on which one can asses 
approximate ones~\cite{Helbig2010}.

Among the unexpected physical traits it can be mentioned the closeness between 
the 
occupation numbers of systems formed by bosons or fermions in the appropriate 
regime \cite{Osenda2015}. In this context, the occupation number of a natural 
orbital refers to the eigenvalue and corresponding eigenvector of a given 
reduced density matrix associated to the quantum state of the system.  We 
identify natural occupation numbers with the eigenvalues of a reduced density 
matrix since they only differ in a constant multiplicative factor: the number of 
particles that constitute the system.

Even for those models with continuous variables where the spectrum, the ground 
state 
and, in some cases, the excited states of a $N$-particle system are exactly 
known, the reduced density matrices that describe the quantum state of a subset 
of $p$ particles are rather difficult to calculate.  

If $N-p$ particles are traced out from the density matrix associated to a 
quantum 
system with $N$ particles, the matrix obtained is usually called a $p$-reduced 
density matrix, or $p$-RDM. The $p$-RDM allows to study a number of physical 
quantities as the natural orbital with its occupation numbers as well as 
different kinds of quantum entropies. Unfortunately, situations where exact 
$p$-RDM can be obtained 
\cite{Osenda2015,Katsura2007,Benavides2014,Schilling2013,kao13} are even more 
scarce than those where exact spectrum or eigenstates are available. The cases 
where a $p$-RDM can be obtained exactly like the Moshinsky \cite{Moshinsky1968}, 
Calogero \cite{Calogero1969} and Calogero-Sutherland \cite{Sutherland1971} 
models, show clearly the difficulties involved. 

The Calogero model, its eigenstates and spectrum, were known to be related to 
many 
other problems in physics, a trademark recognized from the very beginning of the 
subject. In this respect, the pioneering work of Sutherland pointed out that the 
probability distribution of the ground state function for the $N$-particle model 
was identical to the joint probability density function for the eigenvalues of 
random ensembles. In particular, changing the interaction parameter it was 
possible to recover the orthogonal, unitary and symplectic ensembles  density 
functions \cite{Sutherland1971}. This result was first explained as merely 
arising from the Jastrow factor present in the ground state function, but the 
relationship was demonstrated to be deeper than what was originally thought - 
see 
reference \cite{Altshuler1993} where it is shown that the correspondence can be 
extended to response functions or correlations of the density of states of a 
quantum chaotic system \cite{Simons1993}.

At the same time, the relationship of the Calogero model with the fractionary 
quantum Hall effect was well established - see for instance the work by Azuma 
and Iso \cite{Azuma1994}. It was also understood that the Calogero particles are 
basically free but obey generalized fractional exclusion statistics 
\cite{Murthy1994}. So, when referring to bosons or fermions in one- or 
two-dimensional Calogero model, it is the symmetry of the eigenfunctions who 
dictates the terminology since the permutation group in two dimensions allows 
more possibilities to the particles. This arguments explain why the interaction strength 
parameter is sometimes termed ``statistics parameter". 

Following the terminology used by Polychronakos \cite{Poly2006}, the freezing 
trick is the bridge between the Calogero model and lattice integrable systems of the 
Haldane-Shastry type \cite{Poly1993}. It is worth to mention that the trick, 
which is essentially a large interaction strength limit, works well when the 
particles have well defined isolated classical equilibrium positions, as is the 
case of the one-dimensional Calogero model with or without periodic boundary 
conditions. 

Summarizing, the Calogero model has been widely studied from condensed matter 
physics through group theory and has experienced several revivals, that is why 
looking for new physics on it seems always tempting and is rewarding, as we will 
see.

More recently, the availability of exact $p$-RDM \cite{Osenda2015} or very good 
checkable approximations to it \cite{Schilling2013,Benavides2013} has constituted a significant 
tool to shed some light over the behavior of natural occupation numbers in 
fermion systems and their relationship with 
some generalizations of the exclusion principle \cite{Chakraborty2014,Schilling2015}.

The number of non-zero occupation numbers and how fast they become negligible 
are excellent quantifiers to evaluate if an approximate method which involves an 
expansion over a finite functional basis has a good chance to succeed. Of 
course, it is in general impossible to know {\em a priori} how many (if any) 
natural occupation numbers (NONs) become  zero for a given multipartite 
Hamiltonian. Moreover, it is widely accepted that the presence of Coulomb 
``cusps'' leads, inevitably, to an infinite set of non-zero NONs 
\cite{Giesbertz2013}.

Furthermore, the same availability of exact $p$-RDMs develops some unexpected 
features. As has been said above, usually fermion systems have an infinite number 
of nontrivial  NONs. But, as some of us found quite recently, the Calogero model 
in one dimension has a finite number of nonzero NONs for a discrete set of 
values of the interaction parameter \cite{Osenda2015}. To obtain this result it 
is crucial to realize that the $p$-RDM of a system of $N$ particles described by 
the Calogero model can be written exactly as a finite matrix whose entries can 
be obtained analytically. The dimension of the matrix depends on $p$, $N$, the interaction parameter and if the particles are fermions or bosons.

Another feature found in Reference \cite{Osenda2015} is related to the 
behavior of the von Neumann entropy (vNE) obtained from the NONs of 
one-dimensional systems with different number of particles. In all cases, the 
vNE was found to be a non-monotonous function of the interaction strength, 
showing a maximum for some finite value of the interaction strength. 

There are numerous examples which show that different 
entanglement entropies associated to $p$-RDM obtained from ground state 
wave-functions of two- and three-dimensional problems, are monotonous functions 
of the interaction strength between the particles \cite{Manzano,Osenda2007}. 
Moreover, the closely related entanglement properties of fractional quantum Hall 
liquids obtained from the Laughlin wave function also support the monotonous 
behavior. This has been studied in the works by Zeng {\em et al.} 
\cite{Zeng2002}, Iblisdir {\em et al.} \cite{Iblisdir2007} and Haque {\em et 
al.} \cite{Haque2007}. Let us remember that the Laughlin wave function for $n$ 
particles and $1/m$ filling factor has exactly the same form that the ground 
state function of the one-dimensional Calogero model for $n$ particles in one dimension with 
interaction strength $m\,(m-1)$. Anyway, it is prudent not get carried away by the 
similarities, since the partition made to obtain the $p$-RDM will determine between 
which subsystems the entanglement is calculated and a partition between Calogero particles in one- 
or two dimensions is not equivalent to a partition between particles described by 
the Laughlin wave function.

The aim of the present work is to study a few entanglement entropies as functions of the interaction strength for the one and two-dimensional two-particle Calogero model. We will consider a continuous interaction strength parameter, in this way the ground state wave function is exact but the one particle reduced density matrix (1-RDM) and its spectrum are not necessarily so. The large interaction limit will allow us to show that the one-dimensional model has always a finite entanglement entropy in 
contradistinction to the divergent behavior observed in two or larger 
dimensions. In particular, we show that the change from one to two-dimensional behavior can 
be characterized as a crossover, more precisely, the entanglement entropy of 
anisotropic Calogero systems in two dimensions behaves as one-dimensional or two-dimensional accordingly with the amount of anisotropy and the interaction 
strength. It is also shown that the R\'enyi entanglement entropies are able to detect that the system has finite and exact solutions for some particular values of the interaction strength parameter where the effective Hilbert space of the systems is also finite, a fact that is completely 
overlooked by the von Neumann entropy. We also discuss some inadequacy of the so 
called linear entropy to study continuous variable systems in one or two 
dimensions. 

The paper is organized as follows. In Section~\ref{section:the-model} we give some definitions and basic results for the Calogero model and entanglement entropies. In Section~\ref{section:nons}, we calculate the spectrum and von Neumann entropy of the one-dimensional 1-RDM. Section ~\ref{section:renyi} is devoted to the R\'enyi entropies. The two-dimensional isotropic case is studied in Section \ref{section:nons-2d}, while in Section \ref{section:analytical_2D} the anisotropic case is treated in the large interaction limit. We discuss the one to two-dimensional crossover in Section \ref{section:crossover}. Finally, we discuss our findings and conclude in Section~\ref{section:conclusion}.

\section{Preliminaries}\label{section:the-model}

The information content of a given bipartite pure quantum state $|\psi_{AB}\rangle$, can be studied using different entanglement entropies which are obtained from the reduced density matrix $\rho_A = \mbox{Tr}_B (|\psi_{AB}\rangle \langle \psi_{AB}|)$. 

In the case of a two-particle wave function $\Psi(\vec{x}_1,\vec{x}_2)$, where $\vec{x}_1,\,\vec{x}_2$ are the position vectors of the particles, the $1$-RDM can be constructed tracing out one of the particles 

\begin{equation}\label{eq:eprdm}
\rho(\vec{x};\vec{y})
= \int \, \Psi^{\star}(\vec{x},\vec{z}) 
\Psi(\vec{y},\vec{z}) \, d\vec{z} \, .
\end{equation}

Its eigenvalues $\lambda_k$ are given by the following integral equation

\begin{equation}
\label{eprdm}
\int \, \rho(\vec{x};\vec{y}) \, \phi_k(\vec{y}) \, d\vec{y} = \lambda_k 
\phi_k(\vec{x}) \,, \quad k=1,2,3,\ldots \,.
\end{equation}

One of the possible entanglement measure is the von Neumann entropy, $S_{vN}$, which is given by 

\begin{equation}
\label{eq:vNe-def}
S_{vN}(\rho) = - \mbox{Tr} \left(\rho \log_2 \rho 
\right) = - \sum_k \lambda_k \log_2 \lambda_k \,.
\end{equation}

\noindent It is important to emphasize that it is not the only entanglement measure at our disposal. Another possible tool widely used to study entanglement in many-body or extended systems is the R\'enyi entropy 

\begin{equation}
\label{eq:Re-def}
S^{\alpha}(\rho) = \frac{1}{1-\alpha} \,\log_2 \mbox{Tr} \, \rho^{\alpha} = \frac{1}{1-\alpha}\, \log_2 \left( \sum_k \lambda^{\alpha}_k \right)  \,.
\end{equation}

This entanglement measure find their natural place in information theory as a generalization of several other entropies (Shannon's, collision, etc.) which can be recovered for particular values of the parameter $\alpha$. It is worth to mention that for a given probability distribution the R\'enyi entropies defined as in Eq.(\ref{eq:Re-def}) constitute a monoparametric family of convex functions for different choices of the parameter $\alpha$.

The study of R\'enyi entanglement entropies has result in a better understanding of the entanglement in one-dimensional gases and spin chains \cite{Calabrese2011a,Calabrese2011b,Alba2010}. There are a number of reasons to use the quantum R\'enyi entropies, the main two are a) the vNE can be obtained as a limiting case when the parameter $\alpha\rightarrow 1$, and, b) the calculation of the R\'enyi entropies for many different values of the parameter $\alpha$ provides a better understanding of the distribution of the entanglement spectrum of a system than the one obtained by considering only the von Neumann entropy. 

Many authors also use the so called linear entropy (LE), $S_{le}$,

\begin{equation}
\label{eq:def_le}
S_{le} = 1 - \mbox{Tr} \, \rho^2, 
\end{equation}

\noindent mainly motivated by its ease of computation: for continuous variable systems the calculation of $\mbox{Tr} \, \rho^2$ is reduced to just an integral. However, there are some reasons to suspect the quality of information provided by the linear entropy. For instance, no matter how entangled or how many particles are considered, in the large interaction limit the linear entropy of the Calogero model always converges to the unity, as in the case of the Moshinsky model \cite{kao13,Manzano}.

\subsection{The Calogero model}

The two-particle Calogero Hamiltonian in dimension $D$ \cite{Calogero1969} can be written as 

\begin{equation}
\label{ehcal}
H =  h(1) + h(2) + \nu (\nu-1) \frac{1}{r_{12}^2} \,,
\end{equation}

\noindent where  $\vec{r}_{12} = \vec{x}_{1} - \vec{x}_{2}$ denotes the relative 
separation between the particles, $\vec{x}_{1}$ and $\vec{x}_{2}$ are the 
positions of the particles, and $\nu(\nu-1)$ denotes the interaction strength as 
introduced by Sutherland \cite{Sutherland1971}. The one-particle 
harmonic Hamiltonians have the following form 

\begin{equation}\label{ehi}
h(i) = -\frac{1}{2} \nabla_i^2 + \frac12 r_i^2 \; ; \quad i=1,2 \,,
\end{equation}

\noindent where units defined by $\hbar=1$, $m=1$, and 
$\omega=1$ are used through the present work.

\subsection*{One dimensional case}

For two bosons the totally symmetric ground-state wave function and energy are given by

\begin{equation}
\label{ewf0}
E\,=\,\left( \nu+1\right) 
\,;\;\;\;\;\;\psi_0^b(x_1,x_2)\,=\,C^{b}_{1,\nu}\,\Delta_{\nu}\, e^{-\frac{1}{2}
\left(x_1^2+ x_2^2\right)}\,,
\end{equation}

\noindent where  $\Delta_{\nu}$ is the  Jastrow factor

\begin{equation}
\label{ejf}
\Delta_{\nu}\,=\, \left|x_1-x_2\right|^{\nu}\,,
\end{equation}

\noindent while for two spinless fermions we have an anti-symmetrical wave function

\begin{equation}
\label{ewf-f}
\psi_0^{f}(x_1,x_2)\,=\,C_{1,\nu}^f\,sign(x_{1}-x_{2})\, 
\Delta_{\nu}\,
e^{-\frac{1}{2}(x_1^2+x_2^2)}\,,
\end{equation}

\noindent where $C_{1,\nu}^b$ and $C_{1,\nu}^f$ are normalization constants~\cite{fw08}. 

It has recently been shown that for the boson (fermion) wave function with $\nu=2 n\,\, (\nu=2n-1),\;n \in \mathbb{N},$ the absolute value in Eq.~(\ref{ejf}) (Eq.~(\ref{ewf-f})) can be ignored and the only integrals needed to find $1$-RDM are Gaussian integrals with even (odd) powers in the Jastrow factor. Moreover, the $1$-RDM Eq.~(\ref{eq:eprdm}) is then a Gaussian function times a multinomial expression of $(x,y)$. In those cases, the general expression for $\rho_N^{(p)}$, which is quite cumbersome to obtain, can be written as a finite sum of Hermite functions \cite{Osenda2015}. 

\subsection*{Two and higher dimensions}

In dimensions higher than two the exact ground state wave function of bosons

\begin{equation}\label{eq:ground-state-boson}
\Psi_0^b = C_{D,\nu}^b|\vec{x}_1 -\vec{x}_2|^{\mu_b} \, 
e^{  -\frac12 \left( r_1^2 + r_2^2   \right)  },
\end{equation}

\noindent and fermions

\begin{equation}\label{eq:ground-state-fermion}
\Psi_0^f =  C_{D,\nu}^f|\vec{x}_1 -\vec{x}_2|^{\mu_f} \, \psi_S \,
e^{  -\frac12 \left( r_1^2 + r_2^2   \right)  },
\end{equation}

\noindent are quite similar to the one-dimensional ones \cite{Khare1998}. In Eqs.~(\ref{eq:ground-state-boson}) and (\ref{eq:ground-state-fermion}), the exponents  $\mu_b$ and $\mu_f$ are functions of the interaction strength and  the dimension $D$, and $\psi_S$ is one of the $2\times 2$ Slater determinants which are the $N=2$ non-interacting fermion ground state wave functions \cite{Khare1998}.

As in the one-dimensional case, the exact ground state wave function for bosons and fermions cannot be obtained for the same set of parameters since

\begin{equation}\label{eq:expb}
 \mu_b = \frac12 \left( \sqrt{(D-2)^2 +4\nu(\nu-1) }-(D-2) \right) \,,
\end{equation}

\noindent and 

\begin{equation}\label{eq:expf}
 \mu_f = \frac12 \left( \sqrt{D^2 +4\nu(\nu-1) }-D \right)\,,
\end{equation}

\noindent are integer numbers for different values of $\nu$.The $\psi_S$ factor ensures the wave function Eq.~(\ref{eq:ground-state-fermion}) to be totally anti-symmetric with respect to interchange of particles. For $D=2$ there are two such determinants that are linearly independent and can be chosen such that they are both eigenfunctions of the angular momentum operator

\begin{equation}
\psi^{\pm}_S = \left\lbrace
\begin{array}{c}
(x_1-x_2) + i(y_1-y_2) \\
(x_1-x_2) - i(y_1-y_2)
\end{array}
\right.
\,\,\;\quad
L_z \psi^{\pm}_S = \pm\, \psi^{\pm}_S\,.
\end{equation}

Ground-state wave functions can be constructed using linear combinations of $\psi^{\pm}_S$, but this does not imply that their corresponding reduced density matrices have the same entanglement entropies, as we will show in the following sections.


\section{Natural occupation numbers and von Neumann entropy: one-dimensional case}\label{section:nons}

The one-dimensional case was thoroughly analyzed in Reference \cite{Osenda2015} for those values of $\nu$ that are compatible with an exact calculation of the $p$-RDM and its eigenvalues {\em i.e.}  for $\nu=2n$ (bosons) and $\nu=2n+1$ (fermions), with $n$ a natural number. 

In the present work we consider $\nu$ as a continuous variable and calculate using the Rayleigh-Ritz variational method, the eigenvalues of the reduced density matrix Eq.~(\ref{eprdm}). How to use the variational method to calculate an approximate spectrum for a reduced density matrix has been described elsewhere - see References \cite{Osenda2007,Osenda2008,Giesbertz2013}. The natural choice of the basis set are the Hermite functions used to obtain the exact eigenvalues of the finite 1-RDM matrix for integer values of $\nu$~\cite{Osenda2015}.

\begin{figure}
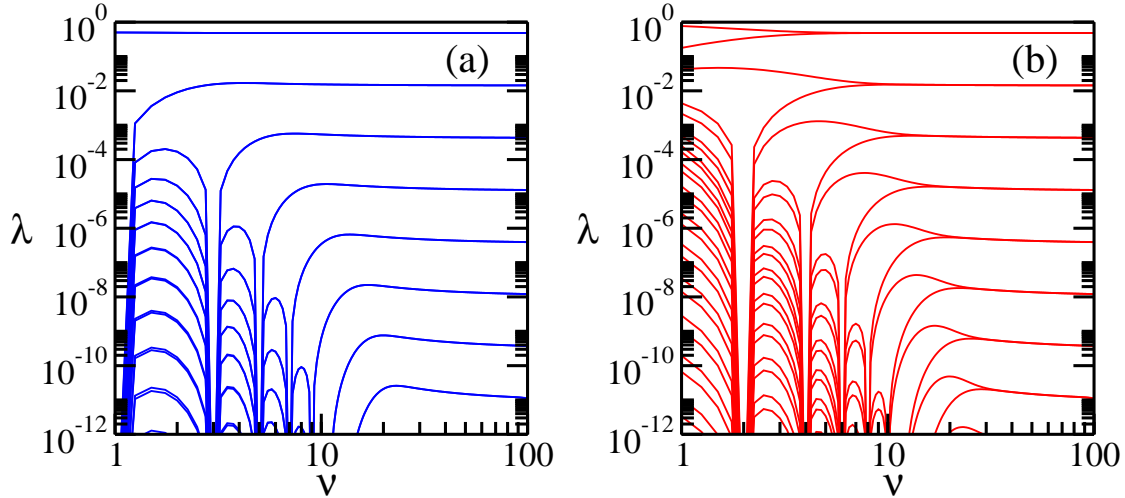

\begin{center}
\includegraphics[width=0.45\textwidth]{fig1a.eps}
\includegraphics[width=0.45\textwidth]{fig1b.eps}
\end{center}
\caption{\label{fig:eigenvalues-1d} Eigenvalues of the 1-RDM for the ground state of a one-dimensional Calogero model from a variational calculation, using 50 one-particle basis functions. Panel (a) shows the results for fermions and (b) those for bosons. The abrupt drop to zero of the eigenvalues at certain integer values (odd for fermions, even for bosons) indicate that the number of natural orbitals is finite. In (a) each eigenvalue is doubly degenerate.}
\end{figure}

The eigenvalues calculated using the variational method for bosons and fermions are shown in a $\log$-$\log$ plot in Fig.~\ref{fig:eigenvalues-1d}. The most salient feature of both sets of curves is the abrupt way in which most eigenvalues drop to zero at the integer values of $\nu$ (see Section~\ref{section:the-model}).  

In the fermion case, since all the eigenvalues are doubly degenerate \cite{Altunbulak2008}, there are only four eigenvalues -the larger ones- that never become null. For $\nu=2n+1$, there are only $2n+2$ non-zero eigenvalues ~\cite{Osenda2015}. The numerical error of the variational eigenvalues for integer values of $\nu$ is $O(\epsilon_m)$ where $\epsilon_m \approx 2\times 10^{-15}$ is the machine precision.

\begin{figure}
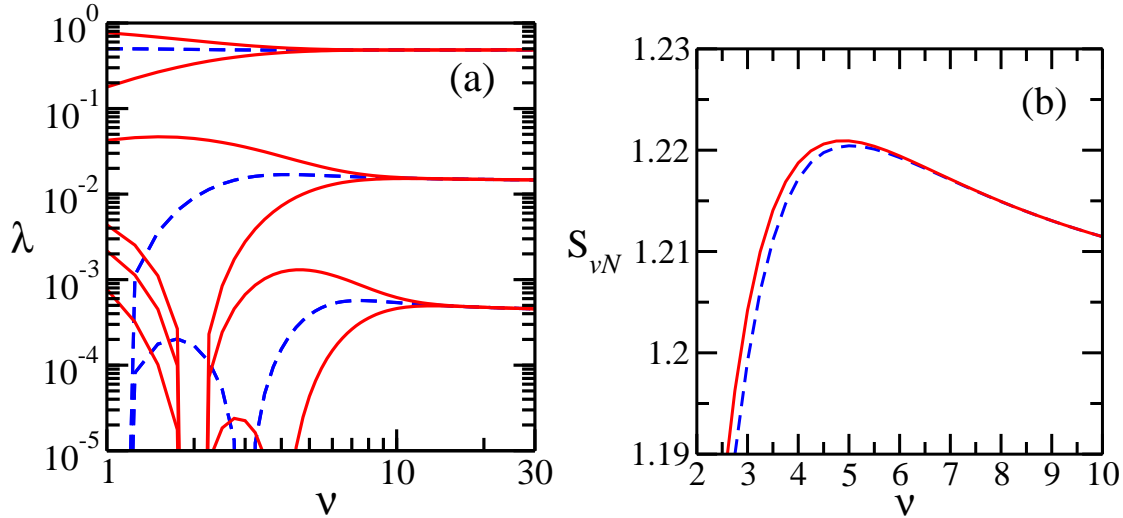

\begin{center}
\includegraphics[width=0.45\textwidth]{fig2a.eps}
\includegraphics[width=0.45\textwidth]{fig2b.eps}
\end{center}
\caption{\label{fig:vonNeumann-1d} (a) Bosons (red line) and fermions (blue dashed line) larger eigenvalues of the 1-RDM and (b) von Neumann entropy for the two particle ground state of a one-dimensional Calogero model from a variational calculation using 50 one-particle basis functions. Note that the eigenvalues of fermions and bosons become degenerate in the large interaction limit.}
\end{figure}

For large values of the interaction parameter $\nu (\nu-1)$, the NONs of bosons and fermions become equal as can be seen in Fig.~\ref{fig:vonNeumann-1d}(a). As a consequence, the von Neumann entropy for both statistics turns out to be the same in the large interaction limit - see Fig.~\ref{fig:vonNeumann-1d}(b). It is important to mention that in this limit, vNE converges to a finite value that can be calculated analytically \cite{Osenda2015,koscik_2015}.

As can be seen in Fig.~\ref{fig:vonNeumann-1d}(b), the vNE shows a maximum around $\nu=5$ for both cases. The appearance of a maximum in the vNE is, at some extent, unexpected since in systems with continuous variables the vNE is known to have a behavior that is strongly correlated to the derivative of the energy with respect to the interaction parameter (see for example~\cite{Osenda2008,Pont2010}). For bound states, it is observed that the vNE increases when the derivative of the energy with respect to the interaction strength diminishes. In our case the derivative increases monotonically but the vNE is not a monotonic function, in contradistinction to what is observed in three-dimensional atom-like systems.

Usually, a non-monotonous behavior of an information content quantifier, as an entanglement measure or an entropy, is related to changes in the analyticity of the ground state energy as happens in a quantum phase transition \cite{Amico2008}. Another reason might be changes in the relative weight between states with different entanglement as happens when the temperature is varied in some thermal mixes \cite{Osenda2005}. The ground state energy and totally symmetric or antisymmetric wave function Eqs.~(\ref{ewf0}) and ~(\ref{ewf-f}) are analytical with respect to the parameter $\nu$. Moreover, the 1-RDM eigenvalues, which are directly related to the vNE, are shown to be analytical around integer $\nu$ using the variational eigenvalues and Finite Size Scaling for quantum mechanics techniques (see Supplemental Material \citep{supmat}).

\section{R\'enyi entropies and finite support of the reduced density matrices}\label{section:renyi}

The smooth behavior of the von Neumann entropy fails to manifest the structure of 
the 1-RDM spectrum as a function of the strength parameter $\nu$. No relevant features are observed at the isolated values of $\nu$ for which the 1-RDM has only a finite set of 
non-zero eigenvalues and the support of the 1-RDM becomes finite \textit{i.e.} the Hilbert space where the system is described becomes finite. As we will show below, the smooth behavior is imposed by the analyticity of the eigenvalues with $\nu$. 

Nevertheless, the structure of the spectrum can be put in evidence by the R\'enyi entropies, defined in Eq. (\ref{eq:Re-def}). As has been pointed out the R\'enyi entropies allow to probe different regions of the spectrum because changing $\alpha$ assigns different weights to the eigenvalues.   

The eigenvalues  of the 1-RDM are analytical functions of $\nu$ (see Supplemental Material \cite{supmat}). We  develop here the bosons case (the fermion case is similar) for $\nu_n=2n$, where the 1-RDM has only $2 n+1$ non-zero eigenvalues. 

The following results will only rely on the analyticity of the eigenvalues around isolated points in the parameter space where the spectrum is finite. As such, they will be valid for any system having this property. We then assume 

\begin{equation}
\label{eq:aee}
\lambda_i(\nu)\sim \left\{ \begin{array}{lll}
\lambda_i(\nu_n) +    \lambda_i^{(1)} (\nu-\nu_n) \; &\mbox{ if } \;i\le2 n+1\\
\mbox{} & \hspace{4cm} \mbox{ for } \nu \rightarrow \nu_n \\
 \lambda_i^{(2)}(\nu-\nu_n)^{2 k_{i,n}}  
\;\;\;\;\;\;\;&\mbox{ if } i>2 n+1\,,
\end{array} \right.
\end{equation}

\noindent where $\lambda_i^{(1)}, \lambda_i^{(2)}$ are constants, and $k_{i,n}\ge 1$ is an integer. Eq. (\ref{eq:Re-def}) can be written as

\begin{eqnarray}
\label{eq:Re-b}
S^\alpha(\nu)&=&
\frac{1}{1-\alpha} \, \log_2{\left(\sum_{i=1}^{2n+1} \lambda_i^\alpha(\nu)  +
{\sum_{i=2n+2}^\infty  \lambda_i^\alpha(\nu) } \right)} \nonumber \\
\mbox{}&=& 
\frac{1}{1-\alpha} \, \left(\log_2{\left(\sum\limits_{i=1}^{2n+1} 
\lambda_i^\alpha(\nu)\right) } +
\log_2{\left(1+\frac{\sum\limits_{i=2n+2}^\infty  \lambda_i^\alpha(\nu) }{
\sum\limits_{i=1}^{2n+1} \lambda_i^\alpha(\nu)} \right)}\right)   \\
\mbox{}& \underset{\nu\rightarrow \nu_n}{\sim}&
\frac{1}{1-\alpha} \, \left(\log_2{\left(\sum\limits_{i=1}^{2n+1} 
\lambda_i^\alpha(\nu)\right) } +
\frac{\sum\limits_{i=2n+2}^\infty  \lambda_i^\alpha(\nu) }{ \ln{2}\,
\sum\limits_{i=1}^{2n+1} \lambda_i^\alpha(\nu)} \right) \,=\,
S_n^\alpha(\nu)+s_n^\alpha(\nu)\,. \nonumber
\end{eqnarray}

\noindent Note that $S_n^\alpha(\nu_n)=S^\alpha(\nu_n)$, and 
$s_n^\alpha(\nu_n)=0$. We can evaluate the derivative of the R\'enyi entropy at 
$\nu=\nu_n$,

\begin{eqnarray}
\label{eq:dsdnu-asym}
\left.\frac{\partial S^\alpha(\nu) }{\partial \nu}\right|_{\nu=\nu_n}&=&
\left.\frac{\partial S_n^\alpha(\nu) }{\partial \nu}\right|_{\nu=\nu_n} +  
\nonumber \\
\mbox{}&&\frac{\alpha}{\ln{2} \, (1-\alpha)} \, \left(
 \frac{\sum\limits_{i=2n+2}^\infty \lambda_i^{\alpha-1}(\nu)
\partial_\nu \lambda_i(\nu)}{\sum\limits_{i=1}^{2n+1} \lambda_i^\alpha(\nu)}
-\frac{\sum\limits_{i=2n+2}^\infty  \lambda_i^\alpha(\nu) 
\sum\limits_{i=1}^{2n+1} \lambda_i^{\alpha-1}(\nu)
\partial_\nu \lambda_i(\nu)}{\left( \sum\limits_{i=1}^{2n+1} \lambda_i^\alpha(\nu)\right) ^2} 
\right)_{\nu=\nu_n} \,. 
\end{eqnarray}

\noindent The first  term in Eq. (\ref{eq:dsdnu-asym}) is a well-defined 
constant
and the third one is zero. As a result of this,  the derivative is dominated 
by the second term. Using the analytic expansion of the eigenvalues, 
Eq. (\ref{eq:aee}), and assuming that 
$k_m$  is the minimum value of $k_{i,n}$, the 
leading asymptotic behavior of $s_n^\alpha$ is

\begin{equation}
\label{eq:sa}
s_n^\alpha(\nu) \underset{\nu\rightarrow \nu_n}{\sim}  
C_n\,((\nu-\nu_n)^{2 k_m})^{\alpha} \,=\, C_n \,
|\nu-\nu_n|^{\delta k_m}\,,
\end{equation}

\noindent where $\delta=2 \alpha$, which implies that

\begin{equation}
\label{eq:dsdnua}
\frac{\partial s_n^\alpha(\nu) }{\partial \nu}\, \underset{\nu\rightarrow 
\nu_n}{\sim}
\delta k_m C_n\,|\nu-\nu_n|^{\delta k_m-1} \,sign(\nu-\nu_n)\,.
\end{equation}

\noindent This equation gives
 
\begin{equation}
\label{eq:dsdnun}
\left.\frac{\partial S^\alpha(\nu) }{\partial \nu}\right|_{\nu=\nu_n}\,=\,
\left\{ \begin{array}{lrl} \left. \begin{array}{lll}
-sign(C_n ) \times \infty  & \mbox{for } &\nu \rightarrow \nu_n^- \\
sign(C_n) \times \infty  & \mbox{for } &\nu \rightarrow \nu_n^+
 \end{array}  \right\}  
& \mbox{if }\;k_m \delta <1  \\
 \left.\begin{array}{lrl}
\partial_\nu S_n^\alpha(\nu_n)\,- C & \mbox{for } &\nu \rightarrow \nu_n^-\\
\partial_\nu S_n^\alpha(\nu_n)\,+ C  & \mbox{for } &\nu \rightarrow \nu_n^+ 
\end{array}  \right\}
  & \mbox{if }\;k_m \delta=1  \\
\partial_\nu S_n^\alpha(\nu_n)   & \mbox{if }\;k_m \delta\ge 1  \,.
  \end{array} \right.
\end{equation}

\noindent Even tough the derivative of $S^\alpha$ is continuous for 
$\delta \ge 1$, it is straightforward to see from the eigenvalue asymptotics, Eq.\ref{eq:aee}, that the second derivative diverges for $1< k_m \delta < 2$, but it is analytical for $k_m \delta=2$, \textit{i.e} the kink at $k_m\delta=1$ is smoothed until it disappears at $k_m \delta=2$.

All our numerical evidence indicate that in the case of the one-dimensional 
Calogero model $k_m=1$ for all values of $i$ and $n$ (see Supplemental Material \cite{supmat}). The R\'enyi entropy 
will then present critical points with infinite derivative for $\alpha<1/2$, a kink
for $\alpha=1/2$ which continuously disappears for increasing $\alpha$ until the value 1 is reached. 
The vNE which can be obtained as the R\'enyi entropy with $\alpha\rightarrow 1$ is then an analytical function of $\nu$.

\begin{figure}
\begin{center}
\includegraphics[width=0.45\textwidth]{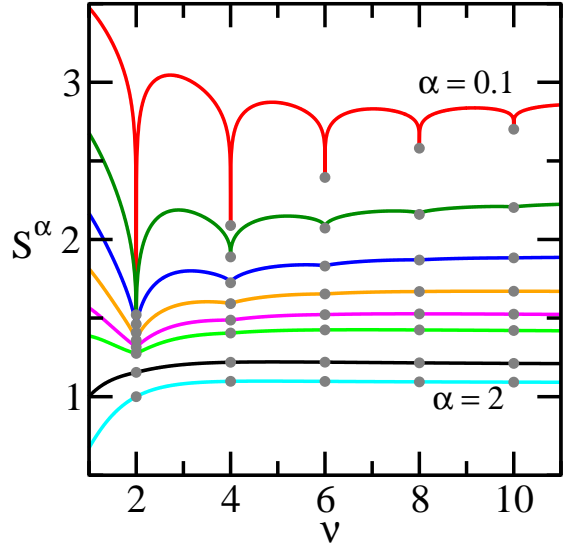}
\end{center}
\caption{\label{fig:renyi1} {One-dimensional bosonic von Neumann entropy (black full line) and R\'enyi entropies as a function of the interaction parameter $\nu$, for $\alpha = 0.1, \,0.2, \,0.3, \,0.4, \,0.5, \,0.6, \,2$ in red, dark green, blue, orange, magenta, light green and cyan full line respectively. The exact values of the entropies for $\nu=2n$ are depicted as gray points.}}
\end{figure}

The non-analytical behavior of the R\'enyi entropies predicted by 
Eq. (\ref{eq:dsdnun}) are a consequence of the eigenvalues analyticity 
assumption, Eq. (\ref{eq:aee}). This salient feature is easily recognizable for $\nu = 2 n$ in Fig. \ref{fig:renyi1}, where the variational R\'enyi entropies for the one-dimensional bosonic Calogero system are shown as a function of the interaction strength parameter for several values of $\alpha$. This figure also shows the exact R\'enyi entropies for those values of $\nu$ for which the 1-RDM has finite support, $\nu=2n$. It is worth to mention that the R\'enyi entropies are decreasing functions of $\alpha$, so the top curve being plotted corresponds to the smaller value chosen for $\alpha$, and that they are bounded from below by the one-dimensional min-entropy $S^\infty_x$ value (see Eq. (\ref{eq:Renyi_2D_ x_limit})).

\begin{figure*}
\begin{center}
\includegraphics[width=0.75\textwidth]{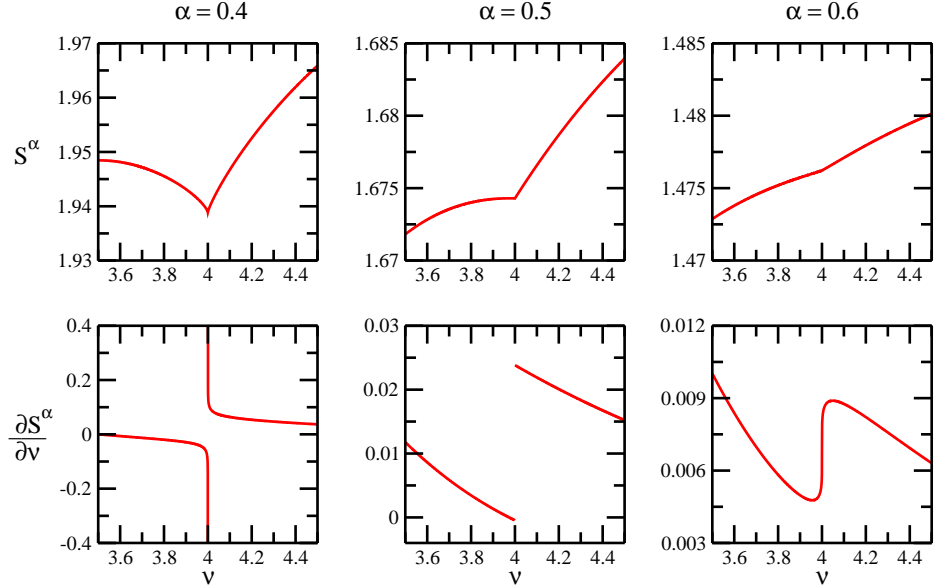}
\end{center}
\caption{\label{fig:renyi2} {One-dimensional bosonic R\'enyi entropies (left column) and their derivatives (right column) as a function of the interaction parameter near $\nu = 4$, for $\alpha = \,0.4, \,0.5, \,0.6$ from top to bottom.}}
\end{figure*}

The R\'enyi entropy and its derivative as a function of the interaction strength parameter near $\nu = 4$ are depicted for $\alpha = 0.4,\,0.5,\,0.6$ in Fig.~\ref{fig:renyi2}. The figure supports all the predictions described above. It shows that the R\'enyi entropy presents a critical point with infinite derivative for $\alpha = 0.4$, a kink with discontinuous derivative for $\alpha =1/2$ and a continuous derivative for $\alpha = 0.6$  with an infinite second derivative.
 
Summarizing, the R\'enyi entropies expose the values of $\nu$ which give a 1-RDM with finite support and this makes them excellent witnesses to detect such a hallmark. Similar features were also seen by Amico and co-workers for $1/2$-spin chains \cite{amico_2013, amico_2013_2, amico_2014, amico_2014_2}. 

\section{Natural occupation numbers and von Neumann entropy: two-dimensional case}\label{section:nons-2d}

\begin{figure}
\begin{center}
\includegraphics[width=0.45\textwidth]{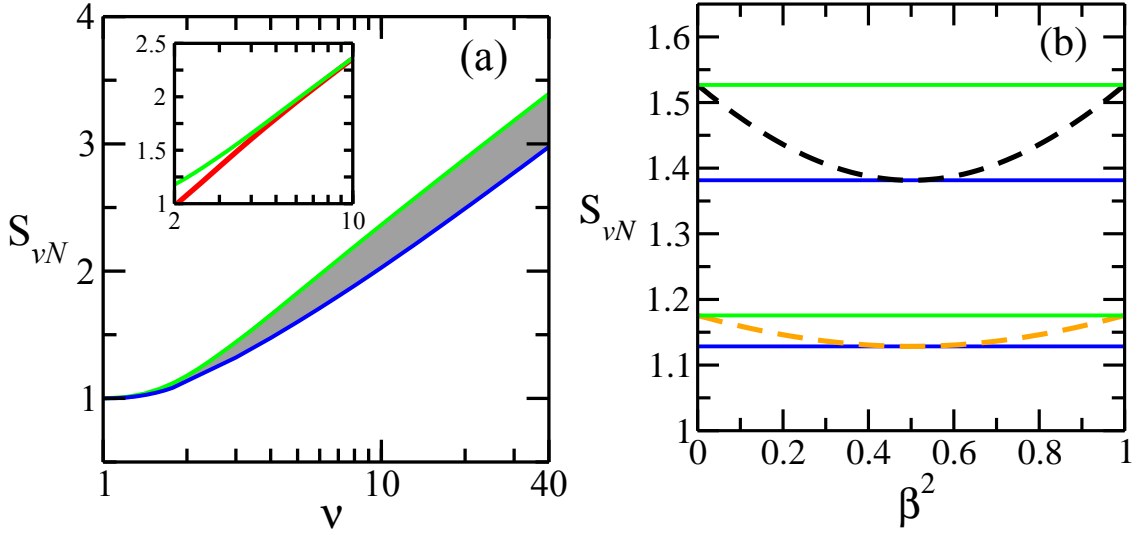}
\includegraphics[width=0.45\textwidth]{fig5b.eps}
\end{center}
\caption{\label{fig:vonNeumann-2d} 
(a) The von Neumann entropy as a function of 
the interaction strength $\nu$ for the rotationally 
invariant fermion wave function $\psi^{+}_S$ (green line) and for the fermion 
wave function $\psi^{x}_S = (x_1 -x_2)$ (blue line). Note the logarithmic 
divergence 
for large interaction strength. Any 
point in the gray shaded area is a ($\nu$, 
$S_{vN}$) 
pair, that can be obtained from a particular choice of the linear combination 
parameter $\beta$ defined in Eq.~(\ref{eq:states-lc}). The inset shows that the 
vNE for the rotationally invariant 
fermion case (green line)  
is asymptotically equal to the vNE for bosons (red line). (b) The fermion 
vNE of $\psi_{lc}(\beta)$ as a function of $\beta^2$ across the shaded region 
for two values of the interacting parameter, $\nu=2$ (orange-dashed line) and 
$\nu=(1+\sqrt{33})/2$ (black-dashed line). The horizontal lines correspond to
the vNE of $\psi^{+}_S$, $\psi^{x}_S$.
}
\end{figure}

The wave function of the two-particle two-dimensional Calogero model is known for all values of $\nu$ (see Eq. (\ref{eq:ground-state-boson}) and (\ref{eq:ground-state-fermion})).  On the other hand, the 1-RDM, its eigenvalues and other 
related quantities are obtained numerically by means of the Rayleigh-Ritz 
variational method. The behavior of the vNE is shown in  Fig.~\ref{fig:vonNeumann-2d}(a) for the boson and fermion cases.  
In this figure the continuous red line corresponds to the bosons case, and the dashed lines to  fermions. The green dashed line corresponds to $\psi_S = \psi_S^+ = (x_1 
-x_2)+ i (y_1 -y_2)$, while the blue dashed line corresponds to $\psi_S = \psi_S^x =\frac{1}{\sqrt{2}}\left(\psi_S^+ + \psi_S^- 
\right)$. 
From the obtained lines in a $\log$ scale it can be seen that the three 
sets of data are consistent with a logarithmic divergence of the vNE when the 
parameter $\nu$ increases, as we will explain in the next section.

With the above definition of $\psi_S^x $ and by defining 
$\psi_S^y =\frac{1}{\sqrt{2}}\left(\psi_S^+ - \psi_S^- \right)$, one can 
construct the related fermion ground state functions $\Psi_0^{f+}$,  
$\Psi_0^{f-}$, $\Psi_0^{fx}$ and $\Psi_0^{fy}$ by substituting  $\psi_S^+$, 
$\psi_S^-$, $\psi_S^x$ and $\psi_S^y$ in Eq.~(\ref{eq:ground-state-fermion}). 
It can then be shown that

\begin{equation}
\label{eq:diferent-psi-nu}
\left| \Psi_0^b \left[\nu(\nu-1)+1\right]\right|^2 = \left| 
\Psi_0^{f\pm}\left[\nu(\nu-1)\right]\right|^2 = \frac12 \left| 
\Psi_0^{fx}\left[\nu(\nu-1)\right]\right|^2 + \frac12 \left| 
\Psi_0^{fy}\left[\nu(\nu-1)\right]\right|^2 ,
\end{equation}

\noindent where the argument between square brackets is the interaction strength 
in Eq.~(\ref{ehcal}) for which the ground state function must be calculated. 

The first two term in the obtained equality ensures that for large enough values 
of  $\nu$ the von Neumann entropies for bosons and  fermions are asymptotically 
the same if the corresponding states are eigenfunctions of $L_z$. So far, we 
have not found  how to translate the relationship between the square modulus of 
the wave functions in Eq.~(\ref{eq:diferent-psi-nu}) to a relationship between 
the NONs of their respective 1-RDM.

The availability of degenerate fermion ground state functions allows us to 
study the von Neumann entropy for different linear combinations of 
orthogonal states. In particular, we studied the vNE of different ground 
state wave functions obtained by replacing $\psi_S$ in
Eq. (\ref{eq:ground-state-fermion}) with the following expression 

\begin{equation}
\label{eq:states-lc}
\psi_{lc}(\beta) = \beta \, \psi_S^+ + \sqrt{1-\beta^2}\; \psi_S^- \,, \quad 
\beta 
\in \left[0,1\right]\,.
\end{equation}

The vNEs for the previously defined 
states,\,$\psi_{lc}(\beta),\,\psi^{x}_S$ and $\Psi_0^b$\,, are shown in 
Fig.~\ref{fig:vonNeumann-2d}. There is a number of interesting features in 
Fig.~\ref{fig:vonNeumann-2d}(a) 
which are noteworthy. The vNEs for all the states defined by 
Eq.~(\ref{eq:states-lc}) diverge logarithmically in the large interaction 
strength limit. For a given $\nu$ value, the entropies of $\psi^{\pm}_S$ 
($\beta=0$, $\beta=1$) are maximal while those of $\psi^{x,y}_S$ 
($\beta=\pm\frac{1}{\sqrt{2}}$) are minimal over $S_{vN}[\psi_{lc}(\beta)]$, 
the shaded region corresponds to all other possible values of 
$\beta$. Also, in the large interaction strength limit, the 
vNEs of both fermions states $\psi_S^{\pm}$ and the bosons ground state are 
asymptotically equal (see the inset in Fig.~\ref{fig:vonNeumann-2d}(a)).  
Figure~\ref{fig:vonNeumann-2d}(b) shows the 
behavior of the vNE calculated for the ground states constructed using 
Eq.~(\ref{eq:states-lc}) as a 
function of $\beta^2$,  for $\nu=2$ (numerical) and for $\mu_f=2 
\Rightarrow\nu=(1+\sqrt{33})/2$ (exact).

The linear entropy shows a different behavior; it 
converges monotonously to unity in the large interaction strength limit. This 
behavior has been reported 
previously for a number of systems and it is usually associated to the competing 
nature of the potentials in the Hamiltonian~\cite{Manzano,koscik_2010}, for 
example in Eq.~(\ref{ehcal}) and (\ref{ehi}) the harmonic term keeps the 
particles near the origin of coordinates, while the repulsive one try to keep 
them as far as possible, especially when $\nu \rightarrow \infty$.

Logarithmic divergences are always elusive to pinpoint when based only in numerical 
data. The reason is that it is quite cumbersome to study large values of $\nu$ 
because of the huge number of base functions needed (that scales as $\nu^2$) to 
put in evidence the logarithmic divergence. Nevertheless, we have been able to 
obtain the vNE for values up to $\nu=80$. To support the numerical evidence 
shown in Fig.~\ref{fig:vonNeumann-2d} we then proceed to study an analytical 
approximation to the two-dimensional problem. 

\section{Analytical treatment of the anisotropic two-dimensional Calogero model in the large interaction strength limit}\label{section:analytical_2D}

The argument stated some paragraphs above about the competing nature of the 
potentials involved in the Calogero model has been useful to develop a method dedicated to 
obtain analytical approximations to the eigenfunctions and eigenvalues of 
Hamiltonians. The method is based on the calculation of the minima of the 
potential and the harmonic approximation consistent with those minima 
~\cite{james_1998,koscik_2015}. Of course for two- or three-dimensional 
problems those minima are not necessarily given by a set of isolated points. 
There is a rather simple way to circumvent the arising difficulties when the 
minima set is not discrete: the potential is ``deformed''  to obtain a finite 
number of minima ~\cite{koscik_2010}. The deformation breaks the symmetry 
between the coordinates. For example in 
the two-dimensional Hamiltonian, we can take $y_i \mapsto \varepsilon y_i$. Within this framework it is possible to study 
the two-dimensional isotropic system as a limiting case of the deformed one, 
therefore, we consider a 
two-dimensional anisotropic Calogero model

\begin{equation}
\label{H_cal_2D_an}
H = -\frac{1}{2}\left( \nabla_1^2+\nabla_2^2\right) + \frac{1}{2}\left\lbrace 
(x_1^2+x_2^2)+\varepsilon^2\,(y_1^2+y_2^2) \right\rbrace + 
\frac{\nu(\nu-1)}{ r_{12}^{2}}\; .
\end{equation}

Introducing the center of mass $\vec{R} = \frac{1}{2} (\vec{x}_1 + 
\vec{x}_2)=(X,Y)$ 
and relative coordinates $\vec{r} = \vec{x}_2-\vec{x}_1=(x,y)$, the Hamiltonian 
(\ref{H_cal_2D_an}) may be written as $H = H^R + H^r$, where

\begin{eqnarray}
\label{H_cal_2D_mc_R}
& &H^R = -\frac{1}{4} \nabla_R^2 + \left(X^2+ \varepsilon^2 \, Y^2\right)\; , \\
\label{H_cal_2D_mc_r}
& &H^r = -\nabla_r^2 + \frac{1}{4}\left( x^2+ \varepsilon^2 \, y^2\right) + 
\frac{\nu(\nu-1)}{\left( x^2+y^2\right)}\; . 
\end{eqnarray}

The wave function is then the product of the center of mass (CM) wave function 
and 
the relative wave function $\Psi(X,Y,x,y) = \psi^R(X,Y) \psi^r(x,y)$ and the 
Schr\"odinger equation separates into two equations

\begin{eqnarray}
\label{Schr_eq_1}
& &H^R\psi^R(\vec{R}) = E^R \psi^R(\vec{R})\; ,\\
\label{Schr_eq_2}
& &H^r\psi^r(\vec{r}) = E^r \psi^r(\vec{r})\; .
\end{eqnarray}

The solutions of the CM Eq. (\ref{Schr_eq_1}) have the following form

\begin{equation}
\label{sol_CM}
\psi_{n,m}^R(\vec{R}) = e^{-X^2}H_n\left( \sqrt{2}X \right) e^{-\varepsilon 
Y^2}H_m\left( \sqrt{2\varepsilon}Y\right)\; ,
\end{equation}

\noindent with energies

\begin{equation}
\label{energy_cm}
E_{n,m}^R = \left(n+\frac{1}{2}\right)+\varepsilon 
\left(m+\frac{1}{2}\right).
\end{equation}

With the aim of solving the relative Schr\"odinger equation 
Eq.~(\ref{Schr_eq_2}) in 
the large interaction strength limit we use the Harmonic Approximation (HA) \cite{james_1998, balzer_2006}. The classical minima of the potential terms are given by $\vec{r}_{min} =\left(\pm \sqrt{2}\,(\nu(\nu-1))^{\frac{1}{4}}, 0 \right)$. In this approximation the Hamiltonian is

\begin{equation}
\label{H_Ha}
H^r_{HA} = -\nabla_r^2 + \frac{1}{2} \left\lbrace 2\,\left( x- x_0  \right)^2 + 
\frac{1}{2} \left(\varepsilon^2 -1 \right) y^2 \right\rbrace\; , 
\end{equation}

\noindent where $x_0 = \sqrt{2}\,(\nu(\nu-1))^{\frac{1}{4}} $. 

The solutions to Eq. (\ref{Schr_eq_2}) are

\begin{equation}
\label{sol_rel}
\psi_{n,m}^r(\vec{r}) = e^{-\frac{(x-x_0)^2}{2}}H_n\left( x-x_0 \right) 
e^{-\frac{\sqrt{\varepsilon^2-1}}{4}\,y^2}H_m\left( 
\left(\frac{\varepsilon^2-1}{4} 
\right)^{1/4} y\right)\; ,
\end{equation}

\noindent with eigenvalues 

\begin{equation}
\label{energy_rel}
E_{n,m}^{r} = 2\, \left(n + \frac{1}{2} \right) + \sqrt{\varepsilon^2 -1} 
\left(m + 
\frac{1}{2} \right)\; .
\end{equation}

From Eq. (\ref{sol_CM}) and (\ref{sol_rel}), the total normalized symmetric 
ground-state wave function can be obtained, 

\begin{equation}
\label{total_wf}
\Psi \left( \vec{r_1}, \vec{r_2} \right) = C \, e^{-\varepsilon \, 
\frac{(y_1+y_2)^2}{4}}           e^{-\sqrt{\varepsilon ^2 -1} \,\frac{ 
(y_2-y_1)^2}{4}} 
e^{-\frac{(\tilde{x}_1+\tilde{x}_2)^2}{4}}        \lbrace    e^{- 
\frac{(\tilde{x}_2-\tilde{x}_1)^2}{2}} +  e^{- 
\frac{(\tilde{x}_1-\tilde{x}_2)^2}{2}}    \rbrace \;,
\end{equation}  

\noindent where $\tilde{x}_1= x_1+\frac{x_0}{2}$ and $\tilde{x}_2= 
x_2-\frac{x_0}{2}$, and 
the normalization constant  

\begin{equation}
\label{normalization}
C = \left(
\frac{
\sqrt{\varepsilon \sqrt{\varepsilon^2-1}}}
{{\sqrt{2} \, \pi^2 \left(  1+e^{- 2 \sqrt{\nu(\nu-1)} } \right) }}
\right)
^{\frac{1}{2}}\; .
\end{equation}

The wave function (\ref{total_wf}) is separable in the $x$ and $y$ coordinates 
and 
can be written as $\Psi\left( \vec{r_1}, \vec{r_2} \right) =  C \, 
\psi_x(\tilde{x}_1,\tilde{x}_2)\, \psi_y(y_1,y_2)$.

Since we are interested in the occupancies of the natural orbitals, we must 
solve 
the integral equation Eq. (\ref{eprdm}). The iterated kernel of a 
symmetric kernel has the same eigenfunctions as the kernel, and the iterated eigenvalues 
are the squared eigenvalues of the kernel \cite{Giesbertz2013,tricomi_1957}, that is, instead of solving Eq. (\ref{eprdm}) one can solve 

\begin{equation}
\label{int_eq}
\int \Psi\left( \vec{r}_1, \vec{r}_2 \right) \phi_k\left(\vec{r}_2 \right) 
d\vec{r}_2 =  \ell_k \, \phi_k\left(\vec{r}_1 \right) \;.
\end{equation}

\noindent The solution to this eigenvalue problem is equivalent to find the Schmidt decomposition of the functions $\psi_x(\tilde{x}_1,\tilde{x}_2)$ and $\psi_y(y_1,y_2)$ given by,

\begin{equation}
\label{psi_x}
\psi_x(\tilde{x}_1,\tilde{x}_2) = 
q(\tilde{x}_1,\tilde{x}_2)+q(\tilde{x}_2,\tilde{x}_1)\; ,
\end{equation}

\noindent where

\begin{equation}
\label{function_q}
q(\tilde{x}_1,\tilde{x}_2) = e^{-\frac{3}{4} 
\left(\tilde{x}_1^2+\tilde{x}_2^2\right) + \frac{1}{2} \tilde{x}_1\tilde{x}_2} 
\; ,
\end{equation}

\noindent and

\begin{equation}
\label{psi_y}
\psi_y(y_1,y_2) = 
e^{-\frac{\varepsilon+\sqrt{\varepsilon^2-1}}{4}\left(y_1^2+y_2^2\right) - 
\frac{\varepsilon-\sqrt{\varepsilon^2-1}}{2}y_1y_2} \; .
\end{equation}

Using the Mehler's formula

\begin{equation}
e^{-(u^2+v^2)\frac{y^2}{1-y^2}+uv \frac{2y}{1-y^2}} = \sum_{k=0}^{\infty} 
\sqrt{1-y^2}\left(\frac{y}{2}\right) \frac{H_k(u) H_k(v)}{k!}\; ,
\end{equation}

\noindent it is possible to find the decomposition of Eq. (\ref{function_q}) 
and (\ref{psi_y}), 

\begin{equation}
\psi(u,v) = \sum_{k=0}^{\infty} \ell_k \, \phi_k(u) \, \phi_k(v)\; .
\end{equation}

After performing some algebra one gets the eigenvalues of the 1-RDM in the limit 
of large interaction strength parameter $\nu$,

\begin{equation}
\label{auval_2D}
\lambda_{k,k^{\prime}} = 2\, \left( 3 \sqrt{2} -4 \right)  \, 
\left(1-\xi(\varepsilon)\right) \xi(\varepsilon)^{k} \, \left( 
17-12\,\sqrt{2} 
\right)^{k^\prime}   \;,
\end{equation}

\noindent where 

\begin{equation}
\label{y_epsilon}
\xi(\varepsilon) = \left( \frac{\left( \varepsilon^2 -1 \right) ^\frac{1}{4} - 
\sqrt{\varepsilon}}{\left( \varepsilon^2 -1 \right) ^\frac{1}{4} + 
\sqrt{\varepsilon}} 
\right)^2
\;.
\end{equation}

Knowing the eigenvalues it is easy to calculate the LE Eq. (\ref{eq:def_le})

\begin{equation}
\label{le_final}
S_{le} = 1 -  \; \left( \frac{3 \sqrt2-4}{9-6 \sqrt 2} \right)  \; \frac{1-\xi(\varepsilon)}{1+\xi(\varepsilon)} \;.
\end{equation}

Since the wave function is separable, the von Neumann entropy presents the form

\begin{equation}
\label{S_VN}
S_{vN} = S^x + S^y(\varepsilon) \;,
\end{equation}

\noindent where each one of the terms in the sum has the form of the one-dimensional vNE \cite{koscik_2015}, {\it i.e.}

\begin{equation}
\label{Sx}
S^x = 1.197371889 \;,
\end{equation}

\begin{equation}
\label{Sy}
S^y(\varepsilon) = - \frac{\ln\left( 
\left(1-\xi(\varepsilon)\right)^{2(1-\xi(\varepsilon))} 
\xi(\varepsilon)^{2\xi(\varepsilon)} \right)}{\ln(4) 
\left(1-\xi(\varepsilon)\right)}\;.
\end{equation}

The R\'enyi entropy Eq. (\ref{eq:Re-def}) in the large interaction limit can be written as

\begin{equation}
\label{eq:Renyi_2D}
S^{\alpha} = S^{\alpha}_{x} + S^{\alpha}_{y} \,,
\end{equation}

\noindent where

\begin{equation}
\label{eq:Renyi_2D_ x}
S^{\alpha}_{x} = \frac{1}{1-\alpha} \log_2 \left(  
\frac{(6 \sqrt{2} - 8)^\alpha}{(1-(17-12\sqrt{2})^\alpha)}
\right) + 1 \,,
\end{equation}

\noindent and

\begin{equation}
\label{eq:Renyi_2D_ y}
S^{\alpha}_{y} = \frac{1}{1-\alpha} \log_2 \left(  
\frac{(1-\xi(\varepsilon))^ \alpha}{(1-\xi(\varepsilon)^\alpha)}
\right) \,.
\end{equation}

The isotropic model can be recovered taking $\varepsilon \rightarrow 1^{+}$. In 
this limit $\xi(\varepsilon) \rightarrow 1$, all the eigenvalues go to zero and the vNE 
diverges logarithmically while the LE goes to one. For any other 
values of $\varepsilon$ the vNE is finite and the LE remains below one. It is 
important to emphasize that the previous analysis can be generalized to dimension $D$ deforming the isotropic potential in $D-1$ dimensions. The one-dimensional problem is recovered for large anisotropy parameter, $\varepsilon >> 1$, case in which  $\xi(\varepsilon) \rightarrow 0$ and consequently 
$S_{le} \rightarrow 1 - \frac{\sqrt{2}}{3}$ and $S_{vN} \rightarrow S^x$.

The R\'enyi entropy as a function of the anisotropy parameter shows the same behavior as the vNE: it diverges logarithmically for $\varepsilon \rightarrow 1^{+}$ and for $\varepsilon >> 1$ it reaches the one-dimensional value $S^\alpha_x$. It is worth to notice that from Eq. 
(\ref{eq:Renyi_2D_ x}) it is straightforward to demonstrate that the one-dimensional 
min-entropy $S^\infty_x$ has the following form

\begin{equation}
\label{eq:Renyi_2D_ x_limit}
S^\infty_x = \lim_{\alpha\to\infty}S^{\alpha}_{x} = \log_2 \left(  
1+\frac{3}{2 \sqrt{2}}
\right) \,.
\end{equation}

\section{Two- to one-dimensional crossover}\label{section:crossover}

As pointed out in section~\ref{section:analytical_2D}, the one-dimensional vNE and LE in the large interaction strength limit are exactly recovered from the two-dimensional model harmonic approximation for large anisotropy parameter. 
This immediately raises the question on how is the two- to one-dimensional \emph{crossover} evidenced in the entropies and whether there is also a similar feature for finite values of the interaction parameter $\nu$. Let us first look at the large interaction limit and then compare it to the numerical results for finite $\nu$.

\begin{figure}
\begin{center}
\includegraphics[height=0.45\textwidth]{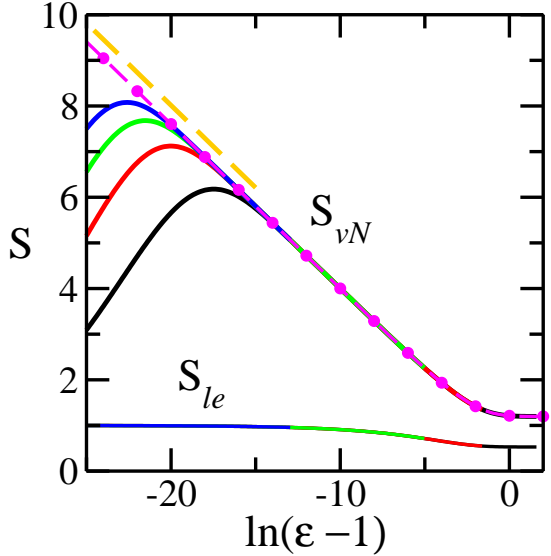}
\end{center}
\caption{\label{fig:s_vn-lim_g_inf} von Neumann (vNE) and linear entropy (LE) 
in the large interaction limit computed using finite sums of the exact 
eigenvalues (full lines) and the exact vNE (magenta dot-dashed line) as a 
function of the anisotropy parameter $\varepsilon$ (Eq.(\ref{S_VN})). The number of  included 
eigenvalues (Eq.(\ref{auval_2D})) are 50 (black), 100 (red), 150 (green) and 200 (blue) respectively. The leading term  of the exact vNE, Eq. (\ref{eq:vNe_limit}), is depicted as a yellow dashed line (shifted to make it visible).}
\end{figure}

The exact vNE of the anisotropic two-dimensional harmonic approximation (Eq. (\ref{S_VN})) is depicted in magenta dot-dashed line in Fig.~\ref{fig:s_vn-lim_g_inf}. Albeit there is not a clear cut criterion to detect the change from two- to one-dimensional behavior, or crossover, one can notice that the one-dimensional limit is reached for $\varepsilon \gtrsim 1.5 $. Moreover, the vNE is finite for any value of $\varepsilon>1$ as it is in the one-dimensional case. 

Calculating the first derivative of Eq. (\ref{Sy}) it is straightforward to demonstrate that

\begin{equation}
\label{eq:vNe_limit}
S_{vN} \sim -\frac{\log(\varepsilon-1)}{\log  
16}  \;\;\;\; \mbox{for}\;\; \varepsilon \sim  1^{+}\,,
\end{equation}

\noindent this behavior is depicted in Fig.~\ref{fig:s_vn-lim_g_inf} as a yellow dashed line which makes the logarithmic divergence of the vNE for $\varepsilon\rightarrow 1^+$ evident.

The entropies obtained adding up the contributions of a finite number of exact eigenvalues (Eq.(\ref{auval_2D})) are also shown in Fig.~\ref{fig:s_vn-lim_g_inf}. 
The plot reveals that no matter how many  eigenvalues are used to evaluate the vNE, there is always a value of $\varepsilon$ for which the vNE reaches a maximum and decays for smaller 
values of the parameter. In other words, the more isotropic 
the system is, the larger number of NONs that are needed to correctly describe the problem. This 
is precisely the reason that makes the identification of a logarithmic 
divergence so difficult, since using a finite numerical approach only provides  
a finite number of approximate eigenvalues to calculate the vNE.

Let us now compare the harmonic approximation to the finite interaction strength results.
In section~\ref{section:nons-2d} we argued that the vNE shown in Fig.~\ref{fig:vonNeumann-2d}(a) grows logarithmically. The divergence for $\varepsilon\rightarrow 1^+$ in the large interaction limit proves that \emph{the vNE of the isotropic Calogero model is infinite} and reinforces what was numerically inferred: the growth is sustained and it is a logarithmic divergence.

More evidence of the convergence of the finite $\nu$ behavior to the one observed in the isotropic harmonic approximation can be obtained studying the ground state energy of the deformed Hamiltonian. The eigenvalues of the relative Hamiltonian, Eq.~(\ref{energy_rel}), depend in a non-analytical fashion on the deformation parameter in the large interaction limit.
We will then compare the ground state energy of the harmonic approximation ($E_{00}^r (\varepsilon)$) to the one obtained from a variational approach in Hamiltonian Eq. (\ref{H_cal_2D_mc_r}) for finite $\nu$ at different anisotropies parameter $\varepsilon$ ($E_{00}^{var}(\nu,\varepsilon)$). We define $E_{00}^{\infty}$ as 

\begin{equation}
\label{eq:variational-relative}
 E_{00}^{\infty} (\nu,\varepsilon) = \frac{E_{00}^{var}(\nu,\varepsilon)}{E_{00}^r(\varepsilon)-1} \,.
\end{equation}

\begin{figure}
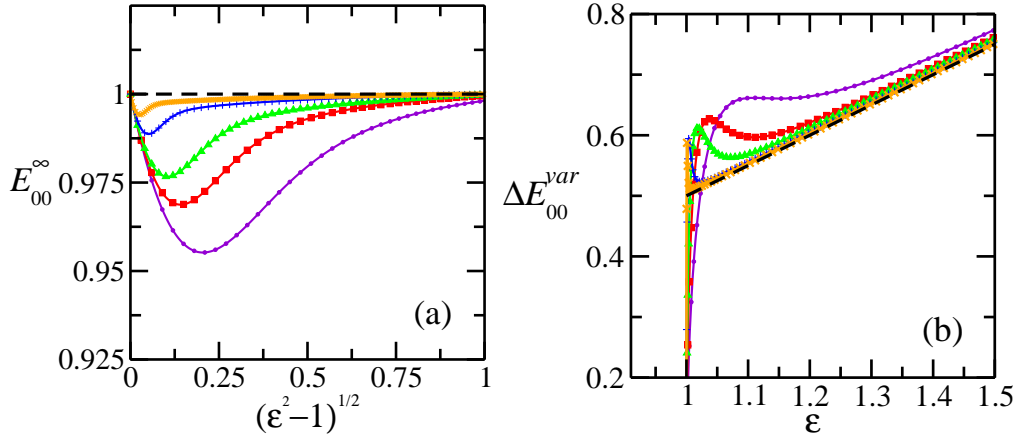

\begin{center}
\includegraphics[height=0.35\textwidth]{fig7a.eps}
\includegraphics[height=0.35\textwidth]{fig7b.eps}
\end{center}
\caption{\label{fig:criticality} (a) The data corresponds to the ratio between 
the 
variational energy and the energy of the relative Hamiltonian in the large 
interaction limit, $E^{\infty}_{00}$ see Eq.~(\ref{eq:variational-relative}). 
The variational energies were calculated for, from bottom to top,
$\nu(\nu-1)=20,50,100,500,2000$. The black dashed line corresponds to the exact 
limit. (b) The numerical derivative of the variational ground state energy times 
$\sqrt{\varepsilon^2-1}$ {\em vs} $\varepsilon$, see 
Eq.~(\ref{eq:asintotico-derivada}) for the precise definition of the function 
$\Delta E_{00}^{var}$. This function was chosen to show the derivative of the ground state energy of the relative Hamiltonian in the limit $\nu\rightarrow 
\infty$ as the black dashed straight line with one-half slope. The other curves 
correspond to the data shown in (a) using the same color convention for the 
different values of $\nu(\nu-1)$.}
\end{figure}

Figure~\ref{fig:criticality}(a) shows how $ E_{00}^{\infty}(\nu,\varepsilon)$ approaches the non-analytical behavior of the function 
$\sqrt{\varepsilon^2-1}$ for large enough values of $\nu$. Notice that for large enough anisotropy $\varepsilon$, the system, no matter how small the interaction strength $\nu$ is, reaches the one-dimensional limit. This observation implies that an anisotropic system should behave as a two-dimensional or one-dimensional one depending on the interplay between the parameters $\nu$ and $\varepsilon$.  

Another quantity that also shows the crossover can be defined as

\begin{equation}
\label{eq:asintotico-diferencia}
\Delta E_{00}^{var}(\varepsilon) = \sqrt{\varepsilon^2 -1} \,\left(
\frac{ E_{00}^{var}(\varepsilon +\Delta \varepsilon) - 
E_{00}^{var}(\varepsilon - \Delta \varepsilon)}{2\,\Delta\varepsilon}
\right) ,
\end{equation}

\noindent and it is displayed in Fig.~\ref{fig:criticality}(b). Due to the dependence of the relative ground state energy, in the  large interaction strength limit, the following relationship is satisfied 

\begin{equation}
\label{eq:asintotico-derivada}
\sqrt{\varepsilon^2-1}\; \frac{dE^r_{00}}{d\varepsilon} = \frac{\varepsilon}{2}\,.
\end{equation}

\noindent Fig.~\ref{fig:criticality}(b) shows how the variational data, 
Eq.~(\ref{eq:asintotico-diferencia}), approaches a straight line with one-half 
slope, which corresponds to the large interaction limit in Eq.~(\ref{eq:asintotico-derivada}).

Summarizing, all the previous analysis indicate a two- to one-dimensional crossover. Moreover, the  vNE diverges logarithmically for the two-dimensional isotropic system, while it remains finite in the anisotropic cases.

\section{Discussion and Conclusions}\label{section:conclusion}

{In the present work, we study the von Neumman and R\'enyi entropies for the two-particle one- and two-dimensional Calogero model. 
We found that the von Neumann entropy of the two-dimensional model with isotropic confinement is a monotonic increasing function of the interaction strength that diverges logarithmically for large interaction strength values, while it remains finite in the anisotropic case as well as in the one-dimensional model. We also show that the one-dimensional behavior is eventually reached when the anisotropy of a two-dimensional system is increased. Using the framework of the harmonic approximation, the crossover from two to one dimensions is demonstrated and it is shown that the von Neumann divergence only occurs in the isotropic case.}

{On the other hand, we found that the R\'enyi entropies expose those values of $\nu$ which give a one-particle reduced density matrix with finite support. Amico and co-workers have found non-analytical behavior for $1/2$-spin chains at the critical values of the Hamiltonian interaction parameters~  \cite{amico_2013, amico_2013_2, amico_2014, amico_2014_2}.}


Let us now discuss the physical implications of the results summarized above. The logarithmic divergence of the von Neumann entropy of the two-dimensional 
Calogero model is, somewhat, to be expected, since the von Neumann entropy of the Laughlin wave function diverges for decreasing filling factors \cite{Iblisdir2007}. However the connection is not direct since the bipartition considered in the work of 
Iblisdir {\em et al.} \cite{Iblisdir2007} differs from the one chosen in the 
present work. Even more, the three-dimensional continuous variable systems studied in the literature support the idea that if the ground state energy of the Hamiltonian  is an analytical and monotonic function of some interaction parameter, so is the von Neumann entropy. These arguments highlight the 
singularity of the one-dimensional case. 

The fact that an anisotropic two-dimensional case behaves like a one-dimensional system in what concerns to its von Neumann entropy, supports the idea that the non-monotonic 
behavior is owed to the restriction of the problem to a ``truncated'' Hilbert 
space. 

In the same sense, the R\'enyi entropies for small enough $\alpha$, have a non-monotonic and non-analytical behavior in the neighborhood of the interaction strength parameter values where the support of the reduced density matrix is finite. It is important to emphasize that \emph{the deduction of the non-analytical behavior of the derivatives of the R\'enyi entropies is completely general}, the only features 
that are unique to the Calogero model are that the values of $\nu$ where the 1-RDM has a finite  entanglement spectrum and the number of non-zero eigenvalues for each one are exactly known.
Consequently, the R\'enyi entropies seems to be a handy tool to detect parameters where a given system possess an exact and finite entanglement spectrum.

The entanglement entropies features commented above are independent of the 
exchange symmetry. Nevertheless, when some particular symmetry is chosen there are several 
aspects that need further discussion. We use bosons and fermions in the sense that the eigenfunctions are symmetrical or anti-symmetrical with respect to coordinate interchange, but in two dimensions the permutation group actually corresponds to the more diverse braid group. 

The eigenvalues of 
the reduced density matrix  for the one-dimensional case for both, bosons and fermions, show two well-defined regimes.
In one regime a given eigenvalue, $\lambda_m$, becomes null for some values of $\nu$, in the other one it is
fairly independent of $\nu$ and $\lambda_m$ seems to obey $\lambda_m\sim a^m$ with $a>0$. Besides,  in the second regime, the natural occupation numbers of bosons and 
fermions have the same asymptotic values in the large interaction limit. Both features, the power law and the asymptotic degeneracy, have been already noted by Schilling in his analysis of the 
one-dimensional harmonium \cite{Schilling2013}.

In two-dimensional models the {natural occupation numbers} of bosons and fermions show the same scenario described in the previous paragraph.
{However, the two-dimensional case presents a fundamental difference with respect to the one-dimensional model because the fermion ground state is twofold degenerate.} So, any function in this two-dimensional functional space, Eq.~(\ref{eq:states-lc}), is a ground state with a particular value of the 
von Neumann entropy. Our results indicate a remarkable physical trait: the fermion states whose 
von Neumann entropy asymptotically approaches the boson's von Neumann entropy are those that are also 
eigenstates of the angular momentum. Moreover, the von Neumann entropy is maximal for these states as shown in Fig.~\ref{fig:vonNeumann-2d}(b).

The analysis of Fig.~\ref{fig:vonNeumann-2d} has led us to think that they can be a particular example 
of a very general result concerning the von Neumann entropy of degenerate states.
We guess that states with more symmetry, as those as $\psi_S^{\pm}$ with 
respect to $\psi_S^{x,y}$ , will always have larger von Neumann entropies than 
those with less symmetry irrespective of the number of particles and particular 
features of the Hamiltonian. More precisely, if $\mathcal{O}$ is an observable which commutates with the Hamiltonian,  $\left[H,\mathcal{O} \right]=0$, and $\psi_{k,l}$, with $l=1, \ldots, L$ are degenerate eigenfunctions of the Hamiltonian 

\begin{equation}
H\psi_{k,l} = E_k \psi_{k,l} ,
\end{equation}

\noindent and eigenfunctions of 

\begin{equation}
 \mathcal{O}\psi_{k,l} = \theta_l \psi_{k,l} ,
\end{equation}

\noindent then $S_{vN}\left[\psi_{k,1}\right] =  \ldots 
=  S_{vN}\left[\psi_{k,L}\right]$ and it is a maximum over 
$S_{vN}\left[\psi\right] 
$ with $\psi\in \mathcal{B}= \mbox{span}\left\lbrace \psi_{k,l} \right\rbrace$. 
Moreover, the minima correspond to the set of equally weighted 
superpositions 

\begin{equation}
\psi_{min} = \frac{1}{\sqrt{L}}\, \sum_{l=1}^L \, e^{i\varphi_l} \psi_{k,l} \, .
\end{equation}

The propositions stated above are valid for all the systems we analyzed using 
numerical methods, prompting us to work on a proof along these lines.

\acknowledgments

We acknowledge SECYT-UNC and CONICET for partial financial support.


\end{document}